\documentclass[11pt]{article}
\usepackage{amsfonts}
\usepackage{amsmath,amsthm}
\usepackage{graphicx,epstopdf,epsfig,multirow,epic,bm}
\usepackage{color}
\usepackage{multicol}
\usepackage{algorithm}
\usepackage{algorithmic}
\usepackage{paralist}

\theoremstyle{definition}
\newtheorem{thm}{Theorem}

\theoremstyle{definition}
\newtheorem{defn}{Definition}
\theoremstyle{definition}

\theoremstyle{definition}

\theoremstyle{definition}

\theoremstyle{definition}

\DeclareGraphicsExtensions{.eps,.eps.gz}
\normalsize \rm
\usepackage{titlesec}

\begin{document}

\begin{center}
{\huge\bf Two Cumulative Distributions For\\[12pt] Scale-freeness of Dynamic Networks}\\[12pt]
\end{center}

\begin{center}
{\large Xiaomin \textsc{Wang}$^{a,b,}$\footnote{Corresponding author, email: wmxwm0616@163.com}
\quad Bing \textsc{Yao}$^{c, d}$}\\[8pt]
{\footnotesize a. Key Laboratory of High-Confidence Software Technologies, Peking University, Beijing 100871, China\\
b. School of Electronics Engineering and Computer Science, Peking University, Beijing,
100871, China\\
c. College of Mathematics and Statistics, Northwest
Normal University, Lanzhou, 730070, China\\
d. School of Electronics and Information Engineering, Lanzhou Jiaotong University, Lanzhou, 730070, China\\[12pt]
}
\end{center}

\begin{quote}
\textbf{Abstract:} It is well-known that the scale-free networks are ubiquitous in nature and society and have been one of the hotspot topic in complex networks. Recently, scholars presented a large quantity of scale-free networks by calculating cumulative distribution. The purpose of this paper is to discuss the relationship between two cumulative distributions, namely, cumulative distribution, edge-cumulative distribution.  Here, firstly, we introduce an relationship between degree distribution and cumulative distribution. Secondly, we introduce the definition of cumulative distribution and edge-cumulative distribution, and compare the relationship between them. Thirdly, we apply algorithmic techniques to construct three deterministic networks, calculate their cumulative distribution and edge-cumulative distribution, and analyze the relationship between cumulative distribution and edge-cumulative distribution. Finally, we offer some open problems for future research in order to understand the interpaly between the degree distribution, cumulative distribution and edge-cumulative distribution. 

\end{quote}

\section{Introduction }

Three physical scientists Newman, Barab\'{a}si and Watts have pointed \cite{Newman-Barabasi-Watts2006}: ``\emph{Pure graph theory is elegant and deep, but it is not especially relevant to networks arising in the real world. Applied graph theory, as its name suggests, is more concerned with real-world network problems, but its approach is oriented toward design and engineering.}'' Graph theory is a fundamental and powerful tool for describing and representing complex networks. It is generally accepted that applied graph theory has an extensive application in complex networks.

Complex networks have been widely considered as an essential and effective instruments for simulating and understanding many real-life networks. There are many examples in our life, such as the World Wide Web \cite{Huberman2001}, food webs \cite{Garlaschelli-2003}, biological networks \cite{Albert-2005}, neural networks \cite{Kim-Lim2015},  information networks \cite{Wu-Shao2015}, co-authorship networks \cite{Barab-Jeong-2002, Pavel-Milos-2013}, and so forth.  Most of complex networks exhibit scale-free property. There are several approaches to verify that a network is scale-free.  
Firstly, Barab\'{a}si and Albert proposed  the \emph{BA-model} in 1999,  to explain the phenomenon of scale-free property by \emph{degree distribution}\cite{Barabasi-Albert1999, Barabasi-Erzsebet2003}.  The degree distribution of BA-model obeys the \emph{power-law}:
$P(k)\propto k^{-\gamma}$ with $2<\gamma<3$. Secondly, Dorogovtsev \emph{et.al} \cite{Dorogovtsev-Goltsev-Mendes-2002} in order to obtain analytical and precise answers for main structural and topological feature of scale-free graphs, defined the \emph{cumulative distribution}. The cumulative distribution of pseudofractal scale-free graphs obeys the \emph{power-law}: $P_{cum}(k)\propto k^{-\gamma+1}$ . Lastly, Liu \emph{et.al} \cite{Liu-Yao-Zhang-2014} put forward a new statistical method, named \emph{edge-cumulative distribution}, which can be applied in determining whether a network is scale-free, i.e,  the edge-cumulative distribution decays as \emph{power-law}: $P_{ecum}(k)\propto k^{-\gamma+1}$.


In the last few decades, some scholars investigated scale-free properties of complex networks by calculating degree distribution, such as,\emph{BA-model} \cite{Barabasi-Albert1999}. While other researchers investigated complex networks by computing cumulative distribution. For instance, Comellas \emph{et.al} \cite{Francesc-2004} discussed the scale-free property of recursive graphs by calculating the cumulative distribution;  Zhang\emph{et.al} \cite{Zhang-Francesc-2005} derived analytical expression of the cumulative distribution of Apollonian networks and explained Apollonian network to be scale-free; Zhang \emph{et.al} \cite{Zhang-zhou-2007} researched the scale-free feature of Sierpinski network model by calculating cumulative distribution. Those works merely discuss the generalization of cumulative distribution of networks, in general. Thus, Liu \emph{et.al} \cite{Liu-Yao-Zhang-2014} explored scale-free properties of complex networks by applying  edge-cumulative distribution. Since then, Wang \emph{et.al} \cite{Wang-Yao-yao2016} studied relationships between degree distribution and cumulative distribution of scale-free networks. Whereas, they did not demonstrate the relationship between the cumulative distribution and  the edge-cumulative distribution. Besides, Wang \emph{et.al} \cite{Wang-Yao-wang2016} proposed newly some mixed cumulative distributions, but they did not mention the difference and the relationship between cumulative distributions. In fact, many literatures applied such statistical approaches to verify scale-free properties of complex networks \cite{Francesc-2004,Zhang-Francesc-2005,Zhang-zhou-2007}.
 Thereby, we focus on finding relationships between cumulative distribution and edge-cumulative distribution by several deterministic networks, as we will show shortly.



The reminder of this paper is organized by the following several sections. 
In Section 2, we introduce the fundamental concepts about the degree distribution and cumulative distribution and discover the relationship between them. In Section 3, we offer the definition of cumulative distribution and edge-cumulative distribution and compare the difference and relationship between them. In Section 4, we show the relationship between cumulative distribution and edge-cumulative distribution for several deterministic networks, such as Recursive graph model, Sierpinski network model, and Apollonian network model. And then, the relationship will help us to estimate the scale-free behavior of particular dynamic networks. Finally, for the outline of this paper, we have to draw a conclusion and bring some open problems for future work in the last section.

For the convenience of discussion and analysis, we introduce some terminologies and notations as follows:
\begin{asparaenum}[$\ast$]
\item $N(t)$ stands for a dynamic network having $n_v(t)$ vertices (nodes) and $n_e(t)$ edges (links) at time step $t$, hereafter, we call $n_v(t)$ the \emph{order} and $n_v(t)$ the \emph{size} of $N(t)$.
\item All vertices of a dynamic network $N(t)$ are collect into the vertex set $V(t)$, and its edges are put into  the edge set $E(t)$.
\item A \emph{complete graph} (\emph{complete network}) of $n$ vertices, denoted as $K_n$, has an edge $uv$ to join each pair of vertices $u$ and $v$. If some subgraph of a graph is a complete graph of $q$ vertices, we call it a \emph{$q$-clique}. A complete graph $K_3$ is called a \emph{triangle}.

\end{asparaenum}

\section{Degree distribution and two cumulative distributions}

In this section, we will introduce the equalvience between the degree distribution and cumulative distribution. 
\subsection{Degree distribution and cumulative distribution}

Barab\'{a}si and Albert \cite{Barabasi-Albert1999} introduced a model with two generic mechanisms, namely, the growth and preferential attachment, and studied its degree distribution. The process can be listed as follows: 










(a) \emph{Growth and preferential attachment}. Add a new vertex $u$ into $N(t-1)$, and join $u$ to vertex $x_i$ of $N(t-1)$ for $i=1,2,\dots, m$ under a preferential attachment $\Pi_i=k_i/\sum_jk_j$; 

(b) \emph{Dynamic equation} (rate equation). Build up a dynamic partial differential equation $\frac{\partial k_i(t)}{\partial t}=m\Pi_i$, and use the initial condition $k_i(t_i)=m$ to solve degree function $k_i(t)$ from the dynamic equation; 

(c) \emph{Degree distribution}. We use a uniformly density function $f(t_i)=(t_i+m_0)^{-1}$ at each time step $t_i$ for computing
\begin{equation}\label{eqa:Barabasi-Albert1999-11}
P(k)=\frac{\partial P(k_i(t)<k)}{\partial k}
\end{equation}
where $k$ is the degree of a vertex, $P(k)$ represents the probability of each vertex in BA-model, then we discover that the degree distribution of BA-model obeys the following equation 
\begin{equation}\label{eqa:Barabasi-Albert1999}
P(k)\sim Ck^{-\gamma}
\end{equation}
where $C$ is a fixed real value. As we observe many real-life networks, one has calculated the degree distribution of many real-life networks by eq.(\ref{eqa:Barabasi-Albert1999}) and discovered that their degree exponent $\gamma$ falls into an interval (2,3]. Then it goes without saying that BA-model is scale-free. After that, many scholars defined some networks and  explained their scale-free properties by computing their degree distribution\cite{Barabasi-Albert1999}. Although many articles apply degree distribution approach, this is not the only approach. There are other approaches to explain scale-free properties of a network.


Differing from degree distribution, Dorogovtsev \emph{et al.} \cite{Dorogovtsev-Goltsev-Mendes-2002} have defined the \emph{cumulative distribution} by
 \begin{equation}\label{cumulative-1}
 P_{cum}(k)=\frac{1}{n_v(t)}\sum_{k'>k}N(k', t)
\end{equation}
for a dynamic network $N(t)$, where $N(k',t)$ is the number of vertices with degree $k'$ in $N(t)$ at time step $t$, $P_{cum}(k)$ represents probability that the degree of the vertex is greater than $k$. Then we discover that the cumulative distribution of network decays the following equation

\begin{equation}\label{cumulative-1}
 P_{cum}(k)\sim k^{1-\gamma}
\end{equation}
Then it can be said with certainty that network $N(t)$ is \emph{scale-free} if its cumulative distribution decays as $P_{cum}(k)\sim k^{1-\gamma}$ with $2<\gamma<3$.

Motivated from  the result obtained in \cite{Dorogovtsev-Mendes-Samukhin2001}, we can calculate
\begin{equation}\label{eqa:degree-distribution}
{
\begin{split}
P(k_i(t)<k)&=1-P(k_i(t)\geq k)=1-\int ^{+\infty }_{k}P(x)dx=1-P_{cum}(k)
\end{split}}
\end{equation}
and then we get the following relationship between the degree distribution and the cumulative distribution as follows
\begin{equation}\label{eqa:degree-distribution-vs-cumulative-distribution}
P(k)=\frac{\partial P(k_i(t)<k)}{\partial k}=-\frac{\partial P_{cum}(k)}{\partial k}\sim (1-\gamma)k^{-\gamma}
\end{equation}

Thus the cumulative distribution $P_{cum}(x)$ also follows power-law distribution, but with a different exponent $1-\gamma$, which is 1 less than the original exponent. Therefore, a network with scale-free behavior is illustrated by deducing that its degree distribution follows eq.(\ref{eqa:Barabasi-Albert1999}) or its cumulative distribution obeys eq.(\ref{cumulative-1}), e.g. BA-model\cite{Barabasi-Albert1999} and pseudofractal graphs\cite{Dorogovtsev-Goltsev-Mendes-2002}. Whereas, the cumulative degree distribution can control the noise problem of statistical data, so the cumulative distribution is generallly used to describe the degree distribution of the network.

\subsection{Two cumulative  distributions}
In this section, we will discuss the relationship between cumulative distribution and edge-cumulative distribution by definition.

According to the definition of \emph{cumulative distribution}, it is given by as follows
 \begin{equation}\label{cumulative}
 P_{cum}(k)=\frac{1}{n_v(t)}\sum_{k'>k}N(k', t)
\end{equation}
for a network $N(t)$, where $N(k',t)$ is the number of vertices with degree $k'$ in $N(t)$ at time step $t$.

Then, Liu \emph{et al.} \cite{Liu-Yao-Zhang-2014} have defined another type of cumulative distribution, called  \emph{edge-cumulative distribution}, as follows
\begin{equation}\label{edge-cumualative}
P_{ecum}(k)=\frac{1}{n_e(t)}\sum_{k'>k}E(k', t)
\end{equation}
for a dynamic network $N(t)$, where  $E(k',t)$ stands for the number of edges incident with the vertices having degrees greater than $k$ in  $N(t)$ at time step $t$, and we call  $N(t)$ a scale-free network as if $P_{ecum}(k)\sim k^{1-\gamma}$ with $2<\gamma<3$.

From the eqs. in (\ref{cumulative}) and (\ref{edge-cumualative}), we can see that the difference between cumulative distribution and edge-cumulative distribution is that the denominator of the expression represents a different implication. $P_{cum}(k)$ and $P_{ecum}(k)$ are applied to the total number of vertices and  the total number of edges of the network, respectively. Besides,  the numerator in a fraction also possess diverse messages in $P_{cum}(k)$ and $P_{ecum}(k)$, the former denotes the number of vertices with degree greater than $k$ at time step $t$; the latter represents the number of edges incident wi{}th the vertices having degrees greater than $k$ at time step $t$.

\section{Connection of two cumulative distributions on deterministic networks}

Mathematical analysis is an effective, powerful tool to investigate the equivalence between two distinct problems. So we can estimate the distance of two-type cumulative distributions by limitation method used in mathematical analysis. Moreover, we will use the so-called algorithmic proof to show three deterministic network models in this article. By the experience and some facts, we conjecture: ``\emph{Two cumulative distributions are equivalent to each other.}'' For the purpose of verifying our conjecture, we introduce three deterministic networks by algorithmic techniques  in the following subsections, and present the proofs for supporting our conjecture.

\subsection{Recursive graph model}

Now we present the  N-algorithm with the \emph{Clique Operation-I} for the  construction of recursive graph  model $K(q,t)$ introduced in \cite{Francesc-2004}, in which scale-free properties of Recursive graph model $K(q,t)$ have been discussed, but not mentioned the two-type cumulative distributions of $K(q,t)$.
{\small
\begin{algorithm}
\caption{\qquad \textbf{N-algorithm}}\label{}
\begin{algorithmic}
\STATE \emph{\textbf{Initialization.}} $K(q,0)$ is a complete graph $K_q$ with $q$ vertices at time step $t=0$.
\STATE \emph{\textbf{Clique Operation-I.}} Add a new vertex  for each of the existing subgraphs being isomorphic to $q$-cliques and join it with all vertices of the subgraph being isomorphic to a $q$-clique .
\STATE \emph{\textbf{Iteration.}}  $K(q,t+1)$ is obtained from $K(q,t)$
by doing a Clique Operation-I to each of the existing subgraphs being isomorphic to $q$-cliques.
\end{algorithmic}
\end{algorithm}
}

Let $\Delta^{N}_{v}(t), \Delta^{N}_{e}(t)$ be the numbers of vertices and edges of Recursive graph model $K(q,t)$ created at time step $t$. According to the N-algorithm,  it is not hard to compute the values of $\Delta^{N}_{v}(t), \Delta^{N}_{e}(t)$ by $\Delta^{N}_{v}(t)=(q+1)^{t-1}, \Delta^{N}_{e}(t)=q(q+1)^{t-1}$. The order $n_v(t)$ and the size $n_e(t)$ represent the total numbers of vertices and edges from $j=0$ to $j=t$, respectively. So we can compute 
\begin{equation}\label{2}
\begin{split}
n_v(t)&=\sum^{t}_{j=0}\Delta^{N}_{v}(j)=q+\sum^{t}\limits_{j=1}(q+1)^{j-1}=\frac{(q+1)^{t}-1}{q}+q, \\
n_e(t)&=\sum^{t}_{j=0}\Delta^{N}_{e}(j)=\frac{q(q-1)}{2}+q\sum^{t}\limits_{j=1}(q+1)^{j-1}=\frac{q(q-1)}{2}+(q+1)^{t}-1
\end{split}
\end{equation}

Futhermore, the number of vertices of degree $d=q,2q,q^{2}+2q,\dots, q+\sum^{t-1}_{j=1}q^{j}$ is  $n_d(t)=(q+1)^{t-1},(q+1)^{t-2},(q+1)^{t-3},\dots, q+1$. Other values of degrees are absent. Therefore, the degree spectrum of $K(q,t)$ is discrete. For large size $n_e(t)$ of $K(q,t)$, we let the vertices of degree $k$ be entered into the network at time step $t=\tau$, so we obtain $k=q(\frac{q^{t-\tau}-1}{q-1}+1)$.

Comellas \emph{et.al} \cite{Francesc-2004} have computed the cumulative distribution  $P^{N}_{cum}(k)$ of Recursive graph model $K(q,t)$ as follows
\begin{equation}\label{eqa:tau-t-000}
{\begin{split}P^{N}_{cum}(k)&=\frac{1}{n_v(t)}\sum^{\tau}_{i=0}\Delta^{N}_{v}(i)
=\frac{(q+1)^{\tau}-1+q^2}{(q+1)^{t}-1+q^2}\propto  (q+1)^{\tau-t}
\end{split}}
\end{equation}
Plugging $\tau-t=\frac{1}{\ln q}\left [\ln\left (\frac{k}{q}-1\right )(q-1)+1\right]$
into (\ref{eqa:tau-t-000}), we obtain
\begin{equation}\label{eqa:tau-t-1}
P^{N}_{cum}(k)\propto k^{1-\gamma_{N}}
\end{equation}
where $\gamma_{N}=1+\frac{\ln(q+1)}{\ln q}$ with $2<\gamma_{N}<3$. Next, we compute the edge-cumulative distribution $P^{N}_{ecum}(k)$ of $K(q,t)$ as below
\begin{equation}\label{eqa:10}
\begin{split}
P^{N}_{ecum}(k)&=\frac{1}{n_e(t)}\sum^{\tau}_{i=0}\Delta^{N}_{e}(i)
=\frac{\frac{q(q+1)}{2}+(q+1)^{\tau}-1}{\frac{q(q+1)}{2}+(q+1)^{t}-1}\propto (q+1)^{\tau-t}
\end{split}
\end{equation}
We can show $|P^{N}_{cum}(k)-P^{N}_{ecum}(k)|<\varepsilon$  for any real number $\varepsilon>0$ (see a detail proof in Appendix A), and claim that both $P^{N}_{cum}(k)$ and  $P^{N}_{ecum}(k)$ are mutually equivalent, and furthermore
$$P^{N}_{ecum}(k)\propto k^{1-\gamma_{N}},~\gamma_{N}=1+\frac{\ln(q+1)}{\ln q}$$

The accurate values of $P^{N}_{cum}(k)$ and $P^{N}_{ecum}(k)$ approximate $(q+1)^{\tau-t}\propto k^{1-\gamma_{N}}$. In other words, $P^{N}_{cum}(k)$ and $P^{N}_{ecum}(k)$ obey the same power-law with the same degree exponent, they  can be used to determine the scale-free properties of $K(q,t)$.
\begin{thm}\label{thm:Recursive-equivalent}
Recursive graph  model $K(q,t)$ holds $P^{N}_{cum}(k)\propto (q+1)^{\tau-t}\propto k^{1-\gamma_{N}}$ and  $P^{N}_{ecum}(k)\propto (q+1)^{\tau-t}\propto k^{1-\gamma_{N}}$ with  $\gamma_{N}=1+\frac{\ln(q+1)}{\ln q}$.
\end{thm}

Also, notice that when $t$ gets in large, the maximal degree of $K(q,t)$ is roughly $q^{t-1}\sim (n_v(t))^{\frac{\ln q}{\ln(q+1)}}=(n_v(t))^{\frac{1}{\gamma_{N}-1}}$.

\subsection{Sierpinski network model}

Sierpinski  network model, denoted as $S(t)$, was introduced in \cite{Zhang-zhou-2007}, but not discussed two-type cumulative distributions. We propose the S-algorithm with the \emph{Fractal Operation} to build up Sierpinski  network model as follows.
{\small
\begin{algorithm}
\caption{\qquad \textbf{S-algorithm}}\label{}
\begin{algorithmic}
\STATE \emph{\textbf{Initialization.}} $S(0)=K_3$. Define a labelling $f$ such that $f(\alpha)=0$ for each vertex $\alpha\in V(0)$, label every vertex $\beta\in V(i)\setminus V(i-1)$ with $f(\beta)=i$.
\STATE \emph{\textbf{Fractal Operation.}} Add a new triangle $\Delta abc$ to each inner triangle $\Delta uvw$  without $f(u)=f(v)=f(w)$,   and then join two ends $a,b$ of the edge $ab$ with $u$, join two ends  $a,c$ of the edge $ac$ with $v$, join two ends $b,c$ of the edge $bc$ with $w$. Figure \ref{fig:fractal-operation} illustrates the process of doing a fractal operation.
\STATE \emph{\textbf{Iteration.}}  $S(t)$ is obtained from the previous $S(t-1)$ by doing a fractal operation to each inner triangle $\Delta xyz$ of $S(t-1)$ without $f(x)=f(y)=f(z)$.
\end{algorithmic}
\end{algorithm}
}

\begin{figure}
\centering
\includegraphics[height=4cm]{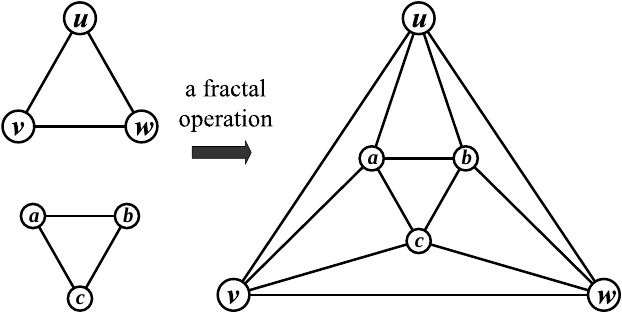}\\
\caption{The process of doing a fractal operation.}
\label{fig:fractal-operation} 
\end{figure}

Two notations $\Delta^{S}_{v}(t), \Delta^{S}_{e}(t)$ denote the numbers of vertices and edges of $S(t)$ created at time step $t$. According to the S-algorithm,  we can calculate the values of $\Delta^{S}_{v}(t), \Delta^{S}_{e}(t)$, in other words, $\Delta^{S}_{v}(t)=3\cdot6^{t-1}, \Delta^{S}_{e}(t)=9\cdot6^{t-1}$. Let $s_v(t), s_e(t)$ be the total  numbers of vertices and edges of the network $S(t)$, from $j=0$ to $j=t$, respectively. Thereby, we have
\begin{equation}\label{3}
\begin{split}
    s_v(t)=\sum^{t}\limits_{j=0}\Delta^{S}_{v}(j)=\frac{3\cdot6^{t}+12}{5},~~
    s_e(t)=\sum^{t}\limits_{j=0}\Delta^{S}_{e}(j)=\frac{9\cdot6^{t}+6}{5}
    \end{split}
\end{equation}
At time $t$, the degree spectrum of $P^{S}_{cum}(k)$ Sierpinski network model $S(t)$ is as: the vertex number $n_d(t)=3\cdot6^{t-1}, 3\cdot6^{t-2}, 3\cdot6^{t-3},\cdots , 6$ when vertex degree $d=4,3^{2}+1,3^{3}+1,\dots, 3^{t}+1$. Other values of degree are absent. We can observe that the degree spectrum of the Sierpinski network model is discrete. The cumulative distribution $P^S_{cum}$ of $S(t)$ is given by
\begin{equation}\label{equ:10}
P^{S}_{cum}(k)=\frac{1}{s_v(t)}\sum^{\tau}\limits_{i=0}\Delta^{S}_v(i)
=\frac{3\cdot6^{\tau}+12}{3\cdot6^{t}+12}\propto 6^{\tau-t}
\end{equation}
Plugging $\tau=t+1-\ln(k-1)/\ln3$ into the eauation  (\ref{equ:10}), we get
\begin{equation}
P^{S}_{cum}(k)\propto6\cdot k^{1-\gamma_{S}}\end{equation}
where $\gamma_S=2+(\ln2/\ln3)$. Next, we  calculate the edge-cumulative distribution of $S(t)$  as follows
\begin{equation}
P^{S}_{ecum}(k)=\frac{1}{s_e(t)}\sum^{\tau}\limits_{i=0}\Delta^{S}_e(i)
=\frac{9\cdot6^{\tau}+6}{9\cdot6^{t}+6}\propto 6^{\tau-t}
\end{equation}
Thereby, $P^{S}_{cum}(k)$ and $P^{S}_{ecum}(k)$ obey the same power-law distribution, that is,
\begin{equation}
P^{S}_{ecum}(k)\propto6\cdot k^{1-\gamma_{S}}
 \end{equation}
with $\gamma_S=2+(\ln2/\ln3)$. It indicates that $|P^{S}_{cum}(k)-P^{S}_{ecum}(k)|<\varepsilon$ (see a proof in Appendix B), and then we report
\begin{thm}\label{thm:Sierpinski-equivalent}
Sierpinski network model $S(t)$ holds $P^{S}_{cum}(k)\sim P^{S}_{ecum}(k)$ with $\gamma_{S}=2+(\ln2/\ln3)$.
\end{thm}
Hence, the edge-cumulative distribution $P^{S}_{ecum}(k)$ of $S(t)$ could be used for determining and explaining $S(t)$ to be scale-free.

\subsection{Apollonian network model}

Apollonian network model has been discussed in \cite{Zhang-Francesc-2005} in detail, except two-type cumulative distributions. For the construction of high-dimensional Apollonian network model, we provide the A-algorithm with the Clique Operation-II as follows.
{\small
\begin{algorithm}
\caption{\qquad \textbf{A-algorithm}}\label{}
\begin{algorithmic}
\STATE \emph{\textbf{Initialization.}} $A(d,0)=K_{d+1}$ with $d\geq 2$.
\STATE \emph{\textbf{Clique Operation-II.}} Add a new vertex to each of the existing subgraphs being isomorphic to $d$-cliques and join it with all vertices of this subgraph being isomorphic a $d$-clique.
\STATE \emph{\textbf{Iteration.}} $A(d,t)$ is obtained from $A(d,t-1)$ by doing a Clique Operation-II to each of the existing subgraphs being isomorphic to $(d+1)$-cliques.
\end{algorithmic}
\end{algorithm}
}

The numbers of vertices and edges of $A(d,t)$ created at time step $t$ are denoted by $\Delta^{A}_{v}(t)$ and $\Delta^{A}_{e}(t)$, respectively. According to the A-algorithm, we have the values of $\Delta^{A}_{v}(t), \Delta^{A}_{e}(t)$, namely, $\Delta^{A}_{v}(t)=(d+1)^{t-1}, \Delta^{A}_{e}(t)=(d+1)^{t}$.
The notations $a_v(t)$ and $a_e(t)$ represent the total number of vertices and the total number of edges of Apollonian network model $A(t)$.
Then it is not difficult to see that the total numbers $a_v(t),a_e(t)$ of vertices and edges of $A(t)$ at time step $t$ can be computed in the following
\begin{equation}\label{4}
\begin{split}
a_v(t)&=\sum^{t}\limits_{j=0}\Delta^{A}_{v}(t)=d+1+\sum^{t}\limits_{j=1}(d+1)^{j-1}=d+1+\frac{(d+1)^{t}-1}{d}\\
a_e(t)&=\sum^{t}\limits_{j=0}\Delta^{A}_{e}(t)=\frac{d(d+1)}{2}+\sum^{t}\limits_{j=1}(d+1)^{j}=\frac{d(d+1)}{2}+\frac{(d+1)^{t+1}-d-1}{d}
\end{split}
\end{equation}
The degree spectrum of the network model is discrete, since the vertex number $n_d(t)=1, d+1, (d+1)^{2}, \dots , (d+1)^{t-1}$ with vertex degree $d=(d+1)(\sum^{t-2}_{j=0}d^{j}+1), (d+1)(\sum^{t-3}_{j=0}d^{j}+1), (d+1)(\sum^{t-4}_{j=0}d^{j}+1), \dots ,d+1$. According to the degree spectrum of Apollonian network model $A(t)$, the cumulative distribution of $A(t)$ is given by
\begin{equation}\label{equ:18}
\begin{split}
P^{A}_{cum}(k)&=\frac{1}{a_v(t)}\sum^{\tau}\limits_{i=0}\Delta^{A}_v(i)=\frac{d(d+1)+(d+1)^{\tau}-1}{d(d+1)+(d+1)^{t}-1}
\propto (d+1)^{\tau-t}
\end{split}
\end{equation}
for large $t$. Moreover, after plugging  $\tau=t+1-\ln(d+1)/\ln d$ into the equation(\ref{equ:18}), we get
\begin{equation}\label{equ:21}
P^{A}_{cum}(k)\propto k^{1-\gamma_{A}}
 \end{equation}
It means that Apollonian network model $A(t)$ is scale-free, because the cumulative distribution follows the power-law with $\gamma_{A}=1+\frac{\ln(d+1)}{\ln d}$.

Now we calculate the edge-cumulative distribution of $A(t)$. It is listed as the following form
\begin{equation}\label{equ:19}
\begin{split}P^{A}_{ecum}(k)&=\frac{1}{a_e(t)}\sum^{\tau}\limits_{i=0}\Delta^{A}_e(i)
=\frac{(d+1)d^2+2[(d+1)^{\tau+1}-d-1]}{(d+1)d^2+2[(d+1)^{t+1}-d-1]}\propto (d+1)^{\tau-t}
\end{split}
\end{equation}
as $t$ becomes large, after plugging  $\tau=t+1-\ln(d+1)/\ln d$  into the above (\ref{equ:19}), we can obtain
\begin{equation}P^{A}_{ecum}(k)\propto k^{1-\gamma_{A}}
  \end{equation}
where $\gamma_{A}=1+\frac{\ln(d+1)}{\ln d}$. On the basis of above analysis and sharply calculation (see a proof in Appendix C), we claim
\begin{thm}\label{thm:Apollonian-equivalent}
Apollonian network model $A(t)$ holds $P^{A}_{cum}(k)\sim P^{A}_{ecum}(k)$.
\end{thm}
The accurate value of the cumulative distribution is related with vertex number, while the edge-cumulative distribution is connected with the edge number. Even though we know the number of vertices and the number of edges are totally different. 
Thereby,  $P^{A}_{cum}(k)$ and $P^{A}_{ecum}(k)$ imply Apollonian network model to be scale-free.

\subsection{Analysis and application}

Three determine network models have distributed us three groups of  cumulative distributions $P^{N}_{cum}(k)$ and $P^{N}_{ecum}(k)$,  $P^{S}_{cum}(k)$ and $P^{S}_{ecum}(k)$, as well as $P^{A}_{cum}(k)$ and $P^{A}_{ecum}(k)$. Clearly, there are
\begin{equation}\label{eqa:c3xxxxx}
P^{N}_{cum}(k)\leq P^{N}_{ecum}(k),~P^{S}_{cum}(k)\leq P^{S}_{ecum}(k),~P^{A}_{cum}(k)\leq P^{A}_{ecum}(k)
\end{equation}

Since $6^{\tau -t}<(d+1)^{\tau -t}\leq (q+1)^{\tau -t}$ with $d\leq q$ according to (\ref{equ:10}), (\ref{equ:18}) and (\ref{eqa:tau-t-000}), we can deduce
\begin{equation}\label{eqa:c3xxxxx}
P^{S}_{cum}(k)<P^{A}_{cum}(k)\leq P^{N}_{cum}(k)
\end{equation}
for larger $t$, Besises, we also have 
\begin{equation}\label{eqa:c3xxxxx}
P^{S}_{ecum}(k)<P^{A}_{ecum}(k)\leq P^{N}_{ecum}(k)
\end{equation}

As application of three groups of  cumulative distributions, by $(\ref{eqa:tau-t-000})$ and $(\ref{equ:19})$, we have
\begin{thm}\label{thm:graceful-vs-odd-graceful}
Suppose that  a dynamic network $L(t)$ with its two cumulative distributions $P^{L}_{cum}(k)$ and $P^{L}_{ecum}(k)$ holds
\begin{equation}\label{eqa:c3xxxxx}
P^{A}_{cum}(k)<P^{L}_{cum}(k)\leq P^{L}_{ecum}(k)\leq P^{N}_{ecum}(k)
\end{equation}
then $P^{L}_{cum}(k)\propto (q+1)^{\tau -t}$ and $P^{L}_{ecum}(k)\propto (q+1)^{\tau -t}$, which mean $L(t)$ obeys the power law.
\end{thm}
Furthermore, $L(t)$ has its own degree distribution $P(k)\propto k^{-\gamma_L}$ with $\gamma_L=1+\frac{\ln (q+1)}{\ln q}$ by Theorem 4. In other words, we need not to compute the exact values of $P^{L}_{cum}(k)$ or $P^{L}_{ecum}(k)$ for deciding $L(t)$ whether takes on scale-free feature.

\begin{defn}\label{defn:maximal-planar}
A maximal planar graph is a planar graph to which no new edges can be added without violating planarity.
\end{defn}

\begin{defn}\label{defn:triangular-synthesized-operation}
Suppose that the bound of outer face of each maximal planar graph $T_i$ is $\{a_i,b_i,c_i\}$ with $i=1,2,3$. We synthesize the edge $a_1c_1$ of $T_1$ with the edge  $b_2c_2$ of $T_2$ into one, and synthesize the edge $a_2c_2$ of $T_2$ with the edge  $b_3c_3$ of $T_3$ into one, and synthesize the edge $a_3c_3$ of $T_3$ with the edge  $b_1c_1$ of $T_1$ into one, the resulting graph is still a maximal planar graph denoted as $G=(T_1\ominus T_2)\ominus T_3$. We call this process a \emph{triangular synthesized operation} on maximal planar graphs. Conversely, doing a \emph{triangular anti-synthesized operation} on $G$ results in three maximal planar graphs $T_1,T_2$ and $T_3$.
\end{defn}

\begin{figure}
\centering{}
\includegraphics[height=3cm]{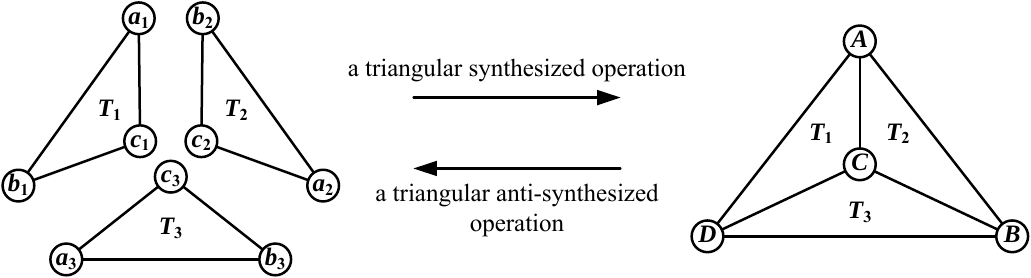}\\
\small{\label{fig:triangular-synthetized-operation}{\footnotesize Figure 2: A scheme for illustrating Definition \ref{defn:triangular-synthesized-operation}.}}
\end{figure}

In Figure 2, we get a maximal planar graph $G=(T_1\ominus T_2)\ominus T_3$ obtained by doing a triangular synthesized operation on three maximal planar graphs $T_1,T_2$ and $T_3$.

\begin{thm}\label{thm:graceful-vs-odd-graceful}
A maximal planar graph $G=(T_1\ominus T_2)\ominus T_3$ is scale-free if each maximal planar graph $T_i$ has its own $P^i_{cum}(k)\sim k^{1-\gamma_i}$ with $i=1,2,3$.
\end{thm}
\textbf{Proof.} By the hypothesis of theorem, each $T_i$ is a maximal planar graph with scale-free feature, the number of vertices of $T_i$ is denoted as $n(T_i)$. The cumulative distribution $P^{T_i}_{cum}(k)$ of $T_i$ obeys $P^{T_i}_{cum}(k)\propto k^{1-\gamma_{i}}$.
\begin{equation}\label{eqa:c3xxxxx}
\begin{split}
P^{T_i}_{cum}(k)&=\frac{1}{n(T_i)}\sum_{k'>k}|V_i(k',t)|=\frac{1}{n(T_i)}\left[n(\tau_{i})\right]\sim k^{1-\gamma_{i}}
\end{split}
\end{equation}
where $n(\tau_{i})$ is the number of vertices with degree $k'$ in $T_i$ at time step $\tau_i$ with $i=1,2,3$.

We do a triangular synthesized operation on maximal planar graph $T_i$, according to the definition of triangular synthesized operation, we can obtain the number $n(G)$ of vertices of the maximal planar graph $G$ is $n(G)=\sum^{3}\limits_{i=1}n(T_i)-5$. Let $V(k',t)$, $V_i(k',t)$ denote as the number of vertices whose degree is greater than $k$ in $G$ and $T_i$ with $i=1,2,3$, respectively. It is generally to compute the cumulative distribution of maximal planar graph
\begin{equation}\label{eqa:c3xxxxx}
\begin{split}
P^{G}_{cum}(k)&=\frac{1}{n(G)}\sum_{k'>k}|V(k',t)|=\frac{1}{n(G)}\left[\sum^3_{i=1}\sum_{k'>k}|V_i(k',t)|\right]\\
&=\frac{1}{n(G)}\left[\sum^3_{i=1}n(\tau_{i})-5\right]\sim\frac{1}{\sum^3\limits_{i=1}n(T_i)-5}\left[\sum^3_{i=1}n(\tau_i)-5\right]\sim k^{1-\gamma_{0}}
\end{split}
\end{equation}
where $\gamma_{0}=\max\{\gamma_{1},\gamma_{2},\gamma_{3}\}$. Therefore, we have proven the conclusion.

Hence, $G=(G_1\ominus G_2)\ominus G_3$ holds $P^{G}_{cum}(k)\sim k^{1-\gamma_{0}}$ if
$P^{G_i}_{cum}(k)\sim k^{1-\gamma{i}}$ with $2<\gamma_i<3$ for $i=1,2,3$, such that $2<\gamma_{0}<3$, $\gamma_{0}=\max\{\gamma_1,\gamma_2,\gamma_3\}$.

\begin{defn}\label{defn:anti-embedded-operation}
Let $G$ be a maximal planar graph, and $\Delta abc$ be an inner face of $G$, and $H$ be another maximal planar graph with the bound $\{a',b',c'\}$ of the out face $\Delta a'b'c'$. We identify the edge $ab$ of the inner face $\Delta abc$ with the edge $a'b'$ of the outer face $\Delta a'b'c'$ of $H$ into one, the edge $bc$ of $\Delta abc$ with the edge $b'c'$ of $\Delta a'b'c'$ into one, and the edge $ca$ of $\Delta abc$ with the edge $c'a'$ of $\Delta a'b'c'$ into one. The resulting graph is still a maximal planar graph, denoted as $G(\ominus H)$, and call the process of producing $G(\ominus H)$ a  \emph{triangular embedded operation}; and the process of splitting $G(\ominus H)$ into two maximal planar graphs $G$ and $H$ is called a  \emph{triangular anti-embedded operation}.
\end{defn}


Figure 3 give an explanation for understanding Definition \ref{defn:anti-embedded-operation}. In Figure 3 an embedded operation is shown from (a) and (b) to (c), and a triangular anti-embedded operation is shown from (c) to  (a) and (b).

\begin{figure}
\centering
\includegraphics[height=3cm]{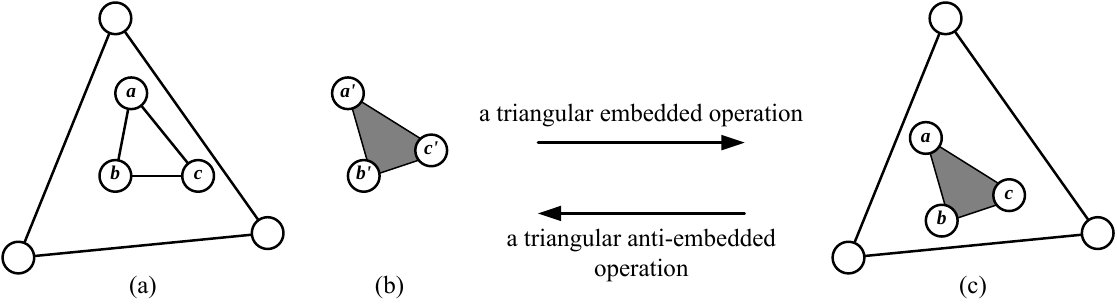}\\
\small{\label{fig:embedded-operation}{\footnotesize Figure 3: (a) A maximal planar graph $G$; (b) another maximal planar graph $H$; (c) the result $G(\ominus H)$ of doing  a triangular embedded operation to $G$.}}
\end{figure}

\begin{thm}\label{thm:graceful-vs-odd-graceful}
A maximal planar graph $G(\ominus H)$ is scale-free if each of  maximal planar graphs $G$ and $H$ are scale-free.
\end{thm}
\textbf{Proof.} By the hypothesis of the theorem, we let the nototation $n(G)$ and $n(H)$ denote the numbers of vertices of $G$ and $H$. Since $G$ and $H$ are scale-free, then we have $G'$s cumulative distribution $P^G_{cum}(k)=\frac{1}{n(G)}|\sum_{k'>k}V(k',t)|\sim k^{1-\gamma_G}$, and $H'$s cumulative distribution $P^H_{cum}(k)=\frac{1}{n(H)}|\sum_{k'>k}V(k',t)|\sim k^{1-\gamma_H}$.

 We do the triangular embedded operation on maximal planar graph $G$ and $H$, according to the definition of triangular embedded operation, the number of $G(\ominus H)$ is given by $G(\ominus H)=n(G)+n(H)-3$, so we can calculate the cumulative distribution $P^{G(\ominus H)}_{cum}(k)$ of $G(\ominus H)$ as follows

 \begin{equation}\label{eqa:c3xxxxx}
\begin{split}
P^{G(\ominus H)}_{cum}(k)&=\frac{1}{n(G(\ominus H))}\sum_{k'>k}|V(k',t)|=\frac{1}{n(G(\ominus H))}\left[n(\tau_G)+n(\tau_H)-3\right]\propto k^{1-\gamma_0}
\end{split}
\end{equation}
where $\gamma_0=\max\{\gamma_G,\gamma_H\}$. Hence, we have completed the proof.

Figure 4 shows us: It is not easy to determine the scale-free properties of maximal planar graphs. By Apollonian network model $A(t)$ and the operations introduce above, we can obtain many random networks $L(t)$ of Theorem 4 with scale-free behaviors.
\begin{figure}
\centering
\includegraphics[height=7cm]{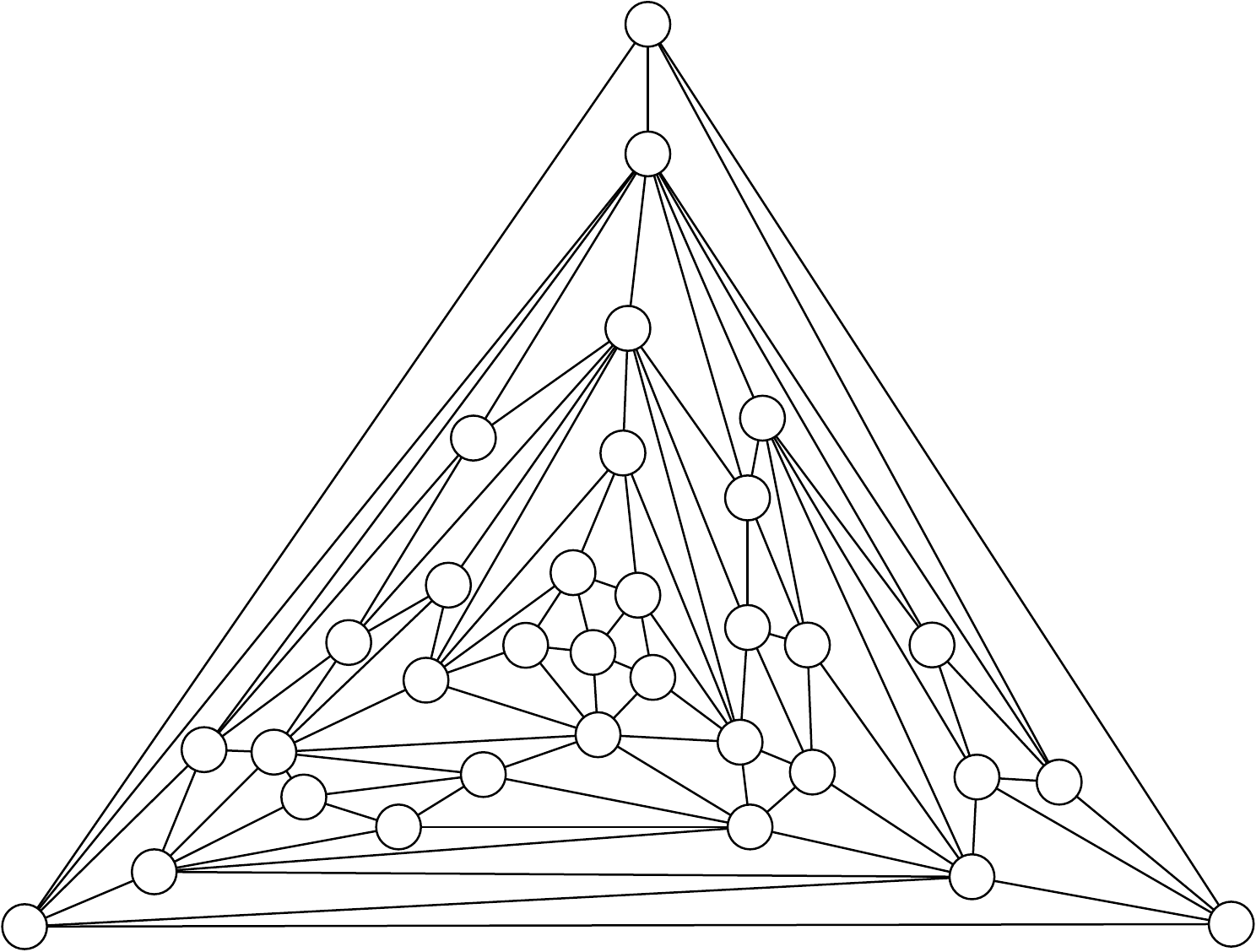}\\
\small{\label{fig:anti-embedded-00}{\footnotesize Figure 4: A maximal planar graph $G^*$ refusing any  triangular anti-embedded operation.}}
\end{figure}

\section{Conclusion}

For a general network $N(t)$, we have shown that the degree distribution $P(k)$ and the cumulative distribution $P_{cum}(k)$ of $N(t)$ are equivalent to each other in (\ref{eqa:degree-distribution-vs-cumulative-distribution}). However, we do not show the equivalence between two cumulative distributions $P_{cum}(k)$ and $P_{ecum}(k)$ for any dynamic network, although we believe it is true.

Since Recursive graph model and Apollonian network model have more edges and high clustering coefficients, we can use them to estimate random network models. For example, Sierpinski network model is really a maximal planar graph, so we want use it to detect scale-free property of some maximal planar graphs. It is a challenge work of finding scale-free maximal planar graphs or scale-free planar graphs.

We analyze  relationships between $P_{cum}(k)$ and $P_{ecum}(k)$, while both of them obey the same power-law distribution at deterministic networks. In order to understand complex networks more deeply, other further researching problems are proposed as follows:

1. We conjecture that cumulative distribution is equivalent to edge-cumulative distribution in deterministic networks. Does this conjecture is available for two-type cumulative distributions of random networks?

2. We did not guarantee whether there exists other approaches for computing  $P_{cum}(k)$ and $P_{ecum}(k)$ which make it become more reasonable. We also find that even though the cumulative distribution and the edge-cumulative  distribution approximate to the same power law exponent, while we can not find an expression to signify equivalence between two cumulative distributions.

3. We do not determine which one is more easier and more efficient and more flexible to be applied in deterministic networks. This problem is required further accurate analysis and numeration and even for need some real problems. We only choose the deterministic scale-free network, this conclusion whether still available for random  networks? 

4. In all deterministic networks or all scale-free deterministic networks, Recursive networks $K(q,t)$ are maximum? 

5. Assume that $N_i(t)$ obeys $P^i_{cum}(k)\sim k^{1-\gamma_i}$, if $2<\gamma_i<\gamma_j<3$, can we guess that  the velocity of $N_i(t)$ networks are growing faster than the velocity of $N_j(t)$ networks? Is there any other methods to determine the velocity of the network?

6. It may be interesting to find a fixed set $F_{tile}$ of maximal planar graphs, such that any maximal planar graph can be tiled by $F_{tile}$, does $F_{tile}$ exist? $F^S_{tile}$ is the set of scale-free maximal planar graphs, so a maximal planar graph $G$ is tiled by the elements of $F^S_{tile}$, then $G$ is scale-free too. By our triangular synthesized operation, triangular embedded operation and triangular anti-embedded operation on planar graphs having triangular outer faces, we can analyze topological structure of a networks, for example, clustering coefficient, shortest diameter, the number of spanning trees, and so on. It is clear that, accomplishment notwithstanding,  scale-free feature of maximal planar graph is a promosing and profound research subject that is merely the begining of a foreseeable far-researching as well as long-sustianable research endeavor.  New discoveries, developments, enhancements and improvements are still yet to come. In the future, we will devote more efforts to invesigate complex networks in order to better help people understand and apply it to explain certain phenomena in real-life. 

\vskip 0.6cm

\noindent \textbf{Acknowledgment.}  
The author, \emph{Bing Yao}, was supported by the National Natural Science Foundation of China under grants No. 61163054, No. 61363060 and No. 61662066.

\renewcommand\refname{References}

\begin{flushleft}
\textbf{Appendix A.} For Recursive graph model $K(q,t)$, we,  by (\ref{eqa:tau-t-000}) and (\ref{eqa:10}), obtain
\end{flushleft}
\begin{equation}\label{eqa:enough-small}
{
\begin{split}
& \quad |P^{N}_{cum}(k)-P^{N}_{ecum}(k)|=\left|\frac{1}{n_v(t)}\sum^{\tau}\limits_{i=0}\Delta^{N}_{v}(i)
-\frac{1}{n_e(t)}\sum^{\tau}\limits_{i=0}\Delta^{N}_{e}(i)\right|\\
&=\left |\frac{(q+1)^{\tau}-1+q^2}{(q+1)^{t}-1+q^2}-\frac{\frac{q(q+1)}{2}+(q+1)^{\tau}-1}{\frac{q(q+1)}{2}+(q+1)^{t}-1}\right |\\
&=\frac{|(q+1)^{\tau}-(q+1)^{t}+2-2q^2|}{[(q+1)^{t}-1+q^2][\frac{q(q+1)}{2}+(q+1)^{t}-1]}\\
&\leq \frac{2(q+1)^{t}+4q^2}{(q+1)^{2t}}\leq \frac{6(q+1)^{t}}{(q+1)^{2t}}=\frac{6}{(q+1)^{t}}<\varepsilon
\end{split}
}
\end{equation}
as $t>\frac{\ln6-\ln\varepsilon}{\ln(q+1)}$, with respect to $t>\tau >0$ and any real number $\varepsilon>0$.

\begin{flushleft}
\textbf{Appendix B.} For Sierpinski network model $S(t)$, we estimate
\end{flushleft}
\begin{equation}
\begin{split}
& \quad  |P^{S}_{cum}(k)-P^{S}_{ecum}(k)|=\left|\frac{1}{s_v(t)}\sum^{\tau}\limits_{i=0}\Delta^{S}_{v}(i)
-\frac{1}{s_e(t)}\sum^{\tau}\limits_{i=0}\Delta^{S}_{e}(i)\right|\\
&=\left|\frac{3\cdot6^{\tau}+12}{3\cdot6^{t}+12}-\frac{9\cdot6^{\tau}+6}{9\cdot6^{t}+6}\right |=\left|\frac{90(6^{t}-6^{\tau})}{(3\cdot6^{t}+12)(9\cdot6^{t}+6)}\right|\\
&\leq\frac{180\cdot6^{t}}{27\cdot6^{2t}}<\frac{8}{6^{t}}<\varepsilon
\end{split}
\end{equation}
to be true for any real number $\varepsilon>0$ and large $t$.

\begin{flushleft}
\textbf{Appendix C.} For Apollonian network model $A(t)$, we verify that both $P^{A}_{cum}(k)$ and $P^{A}_{ecum}(k)$ are mutually equivalent in the following way
\end{flushleft}
\begin{equation}
\begin{split}
& \quad |P^{A}_{cum}(k)-P^{A}_{ecum}(k)|=\left|\frac{1}{a_v(t)}\sum^{\tau}\limits_{i=0}\Delta^{A}_{v}(i)
-\frac{1}{a_e(t)}\sum^{\tau}\limits_{i=0}\Delta^{A}_{e}(i)\right|\\
&=\Biggr |\frac{d(d+1)+(d+1)^{\tau}-1}{d(d+1)+(d+1)^{t}-1}-\frac{(d+1)d^2+2[(d+1)^{\tau+1}-d-1]}{(d+1)d^2+2[(d+1)^{t+1}-d-1]} \Biggr |\\
&\leq \frac{4d(d+1)^{t+2}+2+(2d^{2}+4d)(d+1)^{t+1}}{(d+1)^{2t+1}}\leq\frac{22d^{t+4}}{(d+1)^{2t+1}}<\frac{22}{d^{t-3}}<\varepsilon
\end{split}
\end{equation}
when $t$ gets large, $t>3+\frac{\ln 22-\ln\varepsilon}{\ln d}$, as well as $t>\tau >0$ and arbitrary real number $\varepsilon>0$.

\end{document}